\documentclass[twocolumn,twoside,slac_two]{revtex4}
\usepackage{graphicx}
\usepackage{fancyhdr}
\usepackage{natbib}

\graphicspath{{./figures/}}

\newcommand{\DERIV}[2]{\frac{\partial{}#1}{\partial{}#2}}    

\newcommand{\GJ}[1]{\ensuremath{#1_\textrm{\tiny GJ}}}

\newcommand{\PC}[1]{\ensuremath{#1_\mathrm{pc}}}    

\begin{document}

\title{Self-consistent modeling of pair cascades in the polar cap of a pulsar.}

%

\author{A.~N.~Timokhin} 
\affiliation{Astronomy Department, University
  of California at Berkeley, Berkeley, CA 94720, USA}
\affiliation{Sternberg Astronomical Institute,
  Universitetskij pr. 13, Moscow 119992, Russia}

\begin{abstract}
  Here we briefly report on first results of self-consistent
  simulation of non-stationary electron-positron cascades in the polar
  cap of pulsar using specially developed hybrid PIC/Monte-Carlo
  numerical code.  We consider the case of Ruderman-Sutherland
  cascade -- when particles cannot be extracted from the surface of the
  neutrons star.
\end{abstract}

\maketitle

\thispagestyle{fancy}


\section{INTRODUCTION}

Many previously proposed models for polar cap cascades (and almost all
quantitative models) assumed stationary particle outflow
\citep{Arons1979,Daugherty/Harding82,Muslimov/Tsygan92,
  Harding/Muslimov:heating_2::2002,Hibschman/Arons:pair_multipl::2001}.
Predictions of such models disagree with both observational data
(e.g. the number of electron-positron pairs in pulsar wind nebula is
much higher than predicted, e.g.~\citep{deJager2007}) and results of
numerical models of force-free pulsar magnetosphere (the current
density required to support force-free magnetosphere differs
substantially from what stationary model for polar cap cascade
predicts \citep{Timokhin2006:MNRAS1}).  On the other hand, the
stability of such stationary models has not been quantitatively
studied.

Particle acceleration and electron-positron pair production in the
polar cap could be essentially non-stationary: time intervals of
effective particle acceleration could alternate with intervals when
the accelerating electric field is screened by electron-positron pairs
created in the cap
\citep{Sturrock71,AlBer/Krotova:1975,Ruderman/Sutherland75,Levinson05}.  To
construct a consistent model for particle acceleration in the polar
cap and high energy emission produced there we need to know the
pattern of particle flow.  In our view, the study of electron-positron
cascades should be done starting ab initio.  The key ``ingredients''
of the system must be included in the model: back reaction of
particles on the accelerating electric field and the delay between
photon emission and pair injection.  Possible complexity of the system
behavior compels us to conduct a numerical experiment where particle
acceleration, pair production and variation in the accelerating
electric field are modeled self-consistently.

Here we presents first results of such self-consistent modeling.  We
consider the case of Ruderman-Sutherland cascade and describe the main
properties of the discharge.

\section{NUMERICAL METHOD}

\begin{figure}[b]
  \includegraphics[width=\columnwidth]{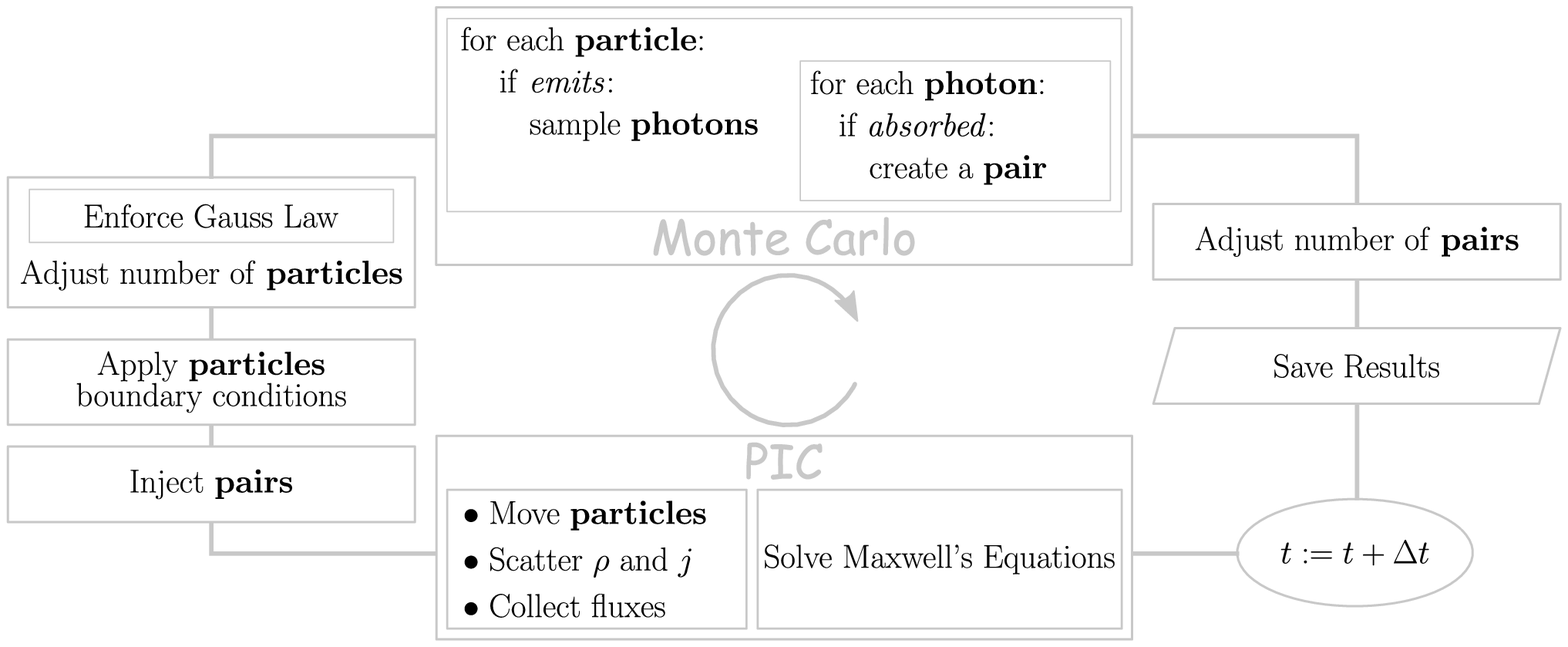}
  \caption{Flow schematic for the hybrid PIC/Monte-Carlo scheme.}
  \label{fig:code}
\end{figure}

For modeling of pair cascades we developed special hybrid PIC/Monte
Carlo code.  Current version of the code is 1D.  The code flow
schematic is shown in Fig.~\ref{fig:code}, the code works as follows.

Plasma dynamics is calculated according to the standard PIC algorithm.
Using the current density known from the previous step we solve
Maxwell equations and get the electric field at the grid points.  Then
for each particle we interpolate the electric field to the particle's
position and get the force on the particle.  By solving the equation
of motion we advance particle's momentum and position.  Motion of
particles across cell boundaries is counted as their contribution to
the electric current which is collected and stored for the next time
step.
 
Photon emission and pair production are calculated as follows. We
sample how many photons capable of producing electron-positron pairs
each particle emits during the current time step. For each emitted
photon we sample its energy from the energy distribution for a given
emission process. Then we sample the distance which the photon will
travel until it gets absorbed. Calculation of the optical depth to
pair creation is done with space steps varying according to the
current value of the cross-section for photon absorption.  Most of the
steps are much larger that the cell size.  The energy of the photon,
the position and the time of the absorption are stored in an array.
At every time step we iterate over that array and pick up photons
which must be absorbed at the time of the current time step. For each
of the selected photons we inject an electron and a positron at the
place of the absorption and delete that photon from the array.  Being
injected at the same point, the electron and the positron do not
change charge and current densities at the moment of injection.

If there are too many numerical particles of some kind in the
computational domain, their number can be reduced by deleting some
randomly selected particles.  Statistical weights of the selected
particles are summed and then the statistical weights of all remaining
particles of the same kind as deleted ones are increased
correspondingly in order to compensate for the deleted particles.

\section{PHYSICAL MODEL}

As previously there were no direct self-consistent kinetic studies of
time depended cascade starting from the first principles, we decided
to address first a simple case in order to develop an intuition of
what kind of plasma behavior to expect and to adjust the numerical
technique accordingly.  \citet{Ruderman/Sutherland75} cascade is the
simplest possible model for the pair cascade in the polar cap -- there
is no plasma inflow from the surface of the neutron star (NS) and all
plasma in the cascade zone is produced by pair creation.

\citet{Ruderman/Sutherland75} estimate for the height of the cascade
zone for young pulsars (their eq.~(22)) gives
\begin{equation}
  \label{eq:h_RS}
  h_{RS}\sim 5\times10^3 \rho_6^{2/7} P^{3/7} B_{12}^{-4/7} \mbox{cm}\,.
\end{equation}
For young pulsars, with periods less than $\sim{}0.1$~sec, $h_{RS}$ is
less that the width of the polar cap
$\PC{r}\simeq{}1.4\times{}10^4/\sqrt{P}$~cm.  Therefore,
one-dimensional approximation should work well for such cascades.  In
1D the continuity equation for the charge density and the Gauss
equation for the accelerating electric field can be combined into the
single equation for the electric field $E$ (see
e.g. \citep{Levinson05})
\begin{equation}
  \label{eq:E}
  \DERIV{E}{t} = -4\pi (j-j_0)\,,
\end{equation}
where $j$ is the actual current density and $j_0$ is the mean current
density flowing through the system.  This is the equation we solve for
the electric field.  It does not need explicitly set boundary
conditions on $E$.  Boundary conditions are implicitly set by the
choice of $j_0$.  The initial electric field distribution is
calculated as solution of 1D Gauss equation for some given initial
charge density distribution and initial boundary conditions.

For young pulsars the dominating emission process in terms of number
of pair-production capable photons is the curvature radiation
(e.g. \citep{Hibschman/Arons:pair_multipl::2001}).  We are primarily
interested in the dynamics of the discharge zone (the region with the
accelerating electric field).  The size of that zone should be of the
order of few $h_{RS}$.  Synchrotron photons emitted by
electron-positron pairs are much less energetic than the curvature
photons and, therefore, these photons have much larger mean free
paths.  They are absorbed at large distances from the NS, where plasma
density is expected to be very high and electric field is already
screened.  Consecutively, pairs produced by the synchrotron photons do
not influence the discharge dynamics and we ignore synchrotron
emission in our simulations.

So, our model for the Ruderman-Sutherland cascade includes 1D
electrodynamics, curvature radiation as the photon emission process,
photon absorption and pair creation in strong magnetic field. Particle
equation of motion includes radiation reaction due to curvature
radiation.

\begin{figure*}[t]

\centering
\includegraphics[width=0.9\textwidth]{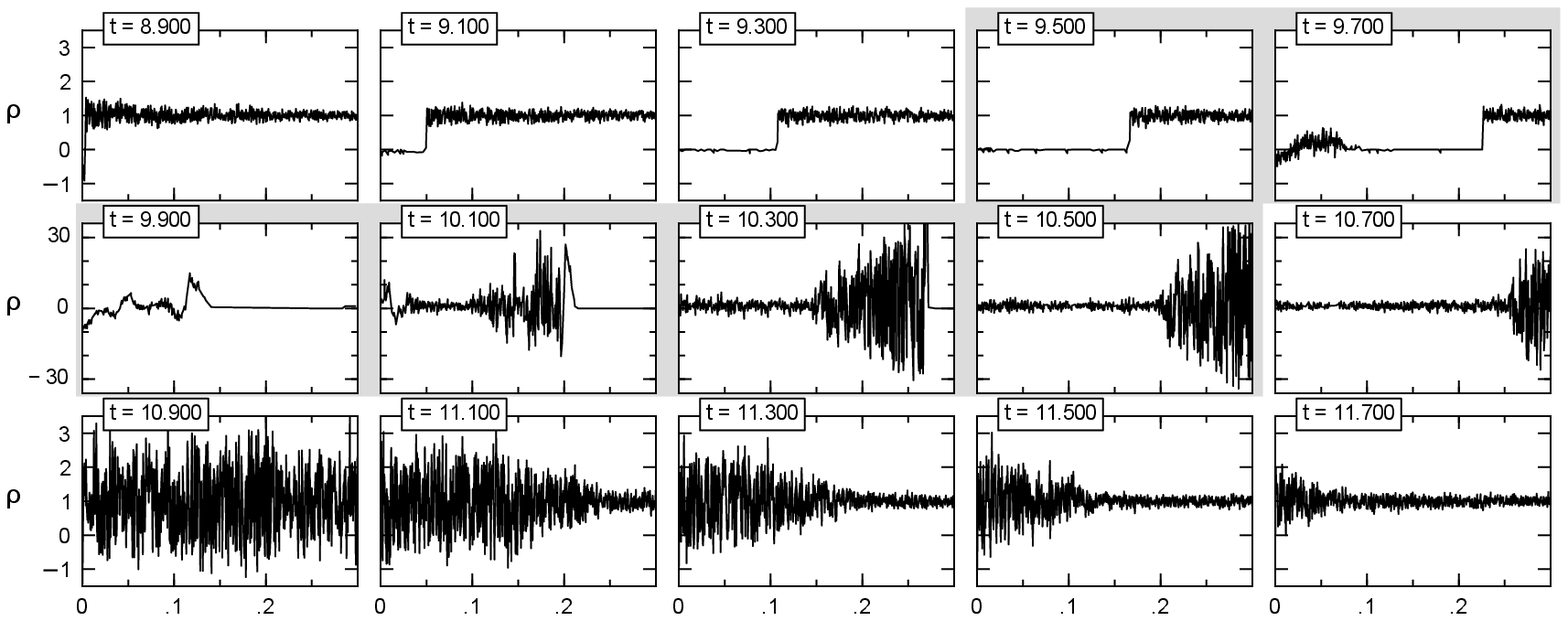}
\includegraphics[width=0.9\textwidth]{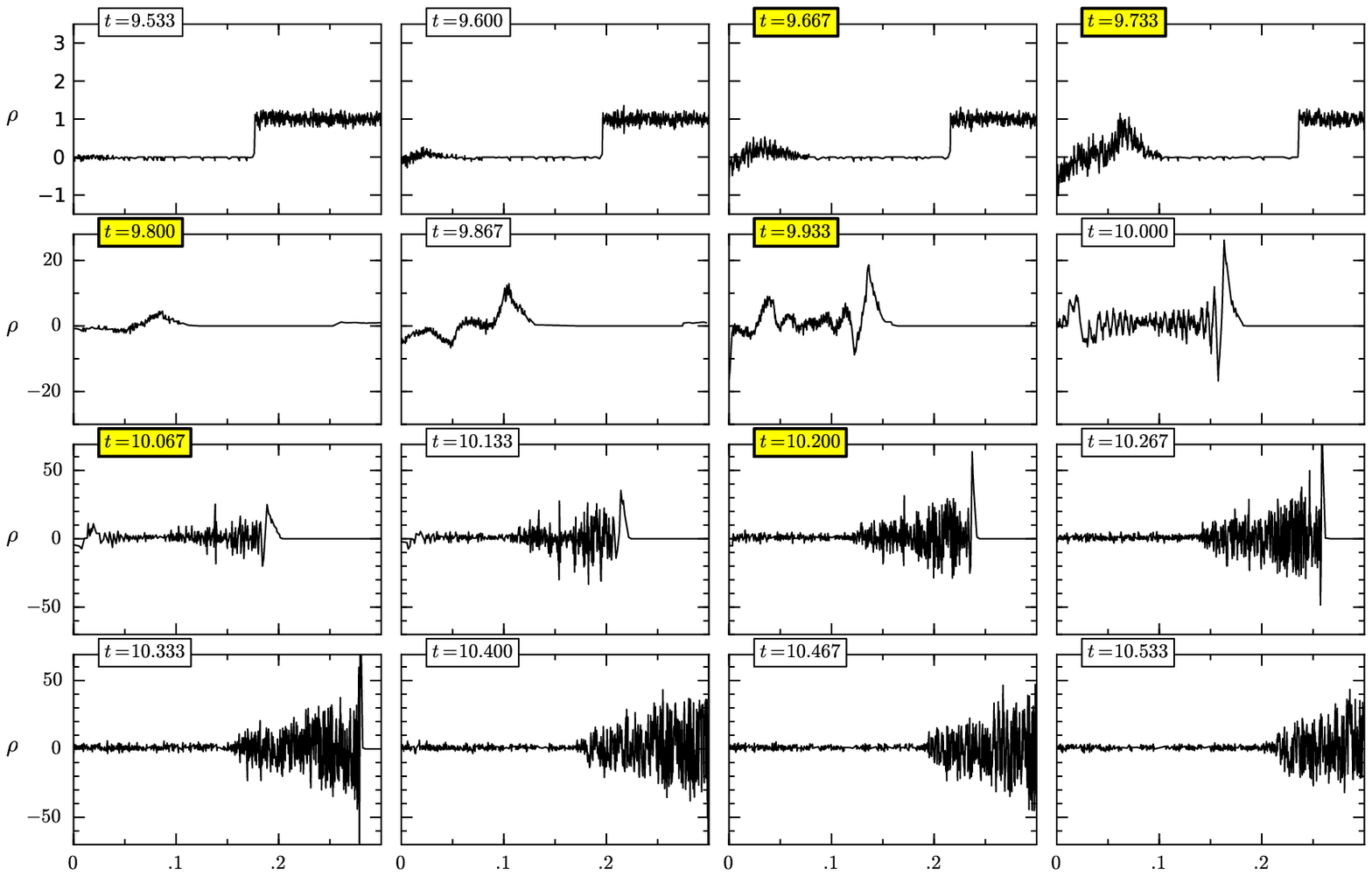}
\caption{Snapshots of a full pair-formation cycle.   
  Charge density is normalized
  to the Goldreich-Julian charge density. Distance is normalized to
  the polar cap radius $\PC{r}$, NS surface is on the left, at
  $x=0$. Pulsar period $P=0.2$~sec, magnetic field $B=10^{12}$~G,
  radius of curvature of the magnetic field lines
  $\rho_{\mbox{cur}}=10^6$~cm. Time, shown in a small box in the
  upper left corner of each plot, is normalized to the flyby time of
  the computation domain. The presented cycle is taken from the
  middle of a long simulation. A typical discharge is shown.
  \textbf{top}~The \emph{whole} cycle of cascade development is
  shown at equally separated moments of time;
  \textbf{bottom}~The formation of pair plasma blob, marked by the
  gray area on the top panel, is shown here with smaller time
  intervals between snapshots.\label{fig:tdc_rho_timeseries}}
\end{figure*}

\section{RESULTS OF NUMERICAL SIMULATIONS}

\begin{figure*}[t]
\centering
\includegraphics[width=0.925\textwidth]{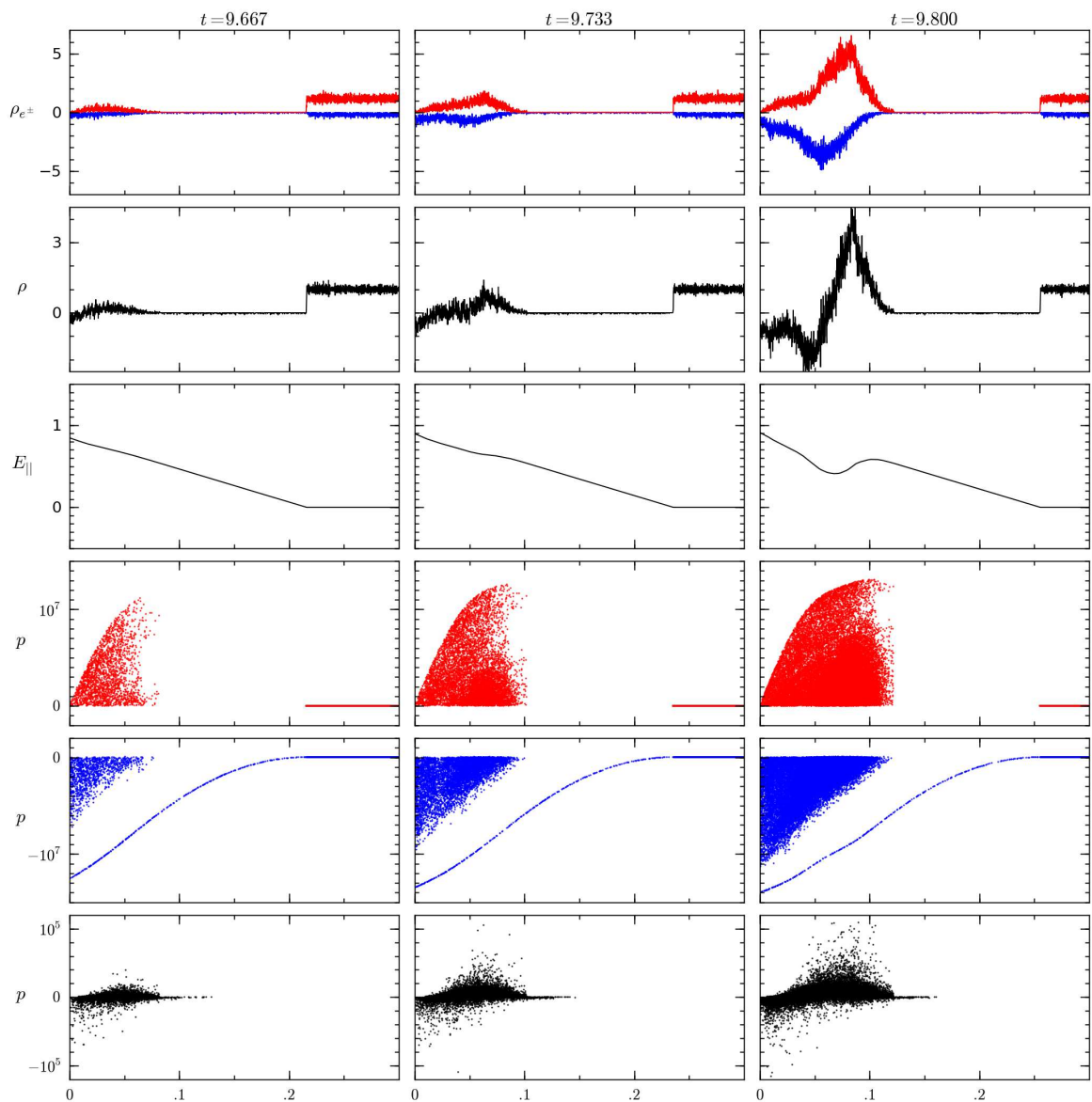}
\caption{Snapshots of cascade development at time steps
    marked on the bottom panel of Fig.~\ref{fig:tdc_rho_timeseries} by
    the yellow time boxes.  At each moment of time several
    characteristic of the cascade zone are shown as functions of the
    distance from the NS:
    \textbf{$\mathbf 1^{st}$ row:} charge density of electrons (blue)
    and positrons (red), $\rho_{e^\pm}$, normalized to the
    Goldreich-Julian charge density;
    \textbf{$\mathbf 2^{nd}$ row:} total charge density $\rho$
    normalized to the Goldreich-Julian charge density;
    \textbf{$\mathbf 3^{rd}$ row:} accelerating electric field
    $E_{||}$ normalized to the vacuum electric field;
    \textbf{$\mathbf 4^{th}$ row:} Phase space of positrons
    (normalized to $m_ec$ momentum $p_{e^+}$ vs coordinate $x$);
    \textbf{$\mathbf 5^{th}$ row:} Phase space of electrons
    (normalized to $m_ec$ momentum $p_{e^-}$ vs coordinate $x$);
    \textbf{$\mathbf 6^{th}$ row:} Phase space of pair-producing
    gamma-rays (normalized to $m_ec$ momentum $p_{\gamma}$ vs
    coordinate $x$).
    \label{fig:tdc_details}} 
\end{figure*}

\begin{figure*}[t]
\centering
\includegraphics[width=0.925\textwidth]{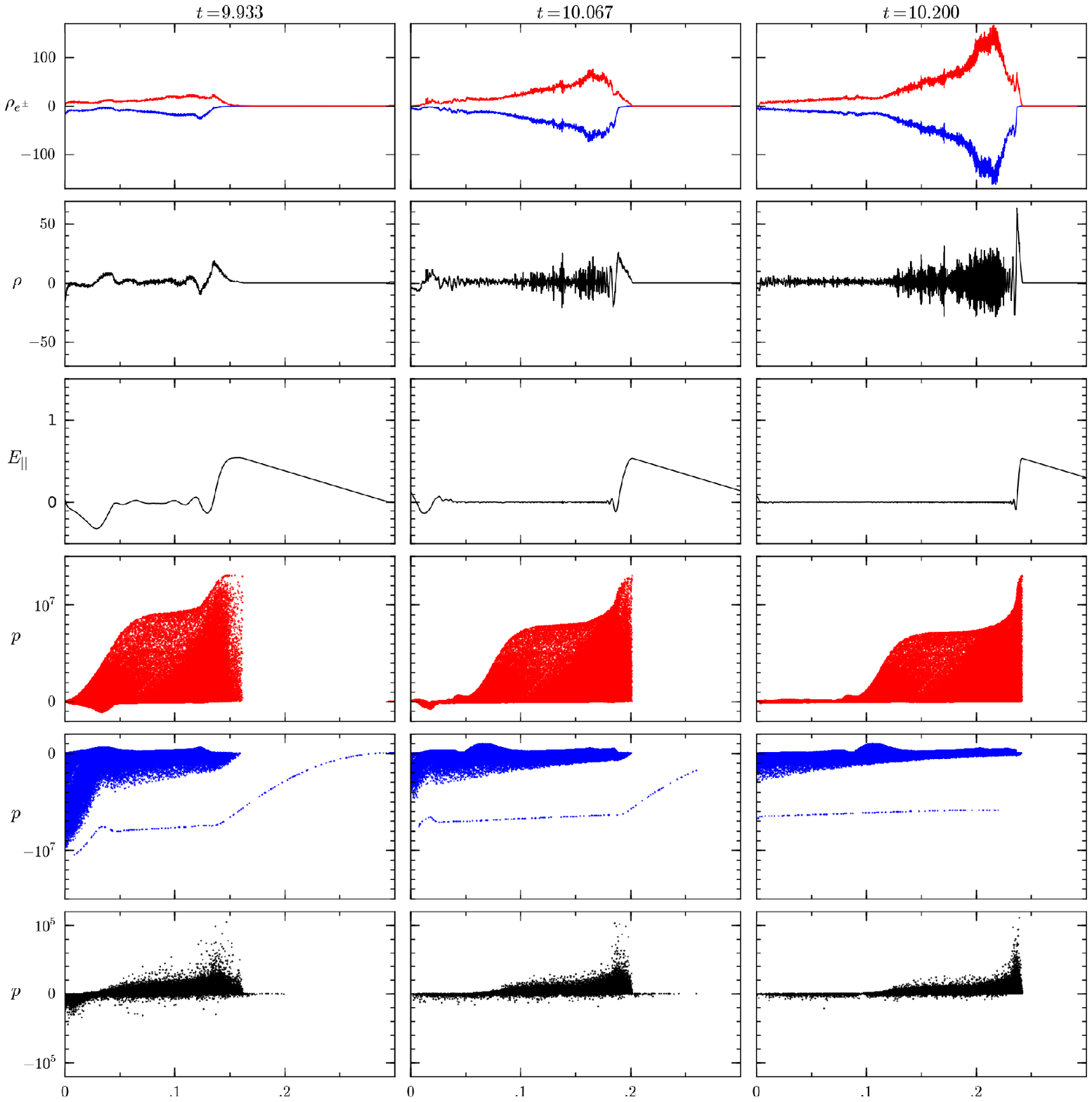}
\caption{Snapshots of cascade development at time steps marked on the
  bottom panel of Fig.~\ref{fig:tdc_rho_timeseries} by the yellow time
  boxes (continued from Fig.~\ref{fig:tdc_details}). Notations are
  the same as in Fig.~\ref{fig:tdc_details}.
  Note the change of y-axis normalization for
  charge densities (the first two rows) compared to Fig.~\ref{fig:tdc_details}.
  \label{fig:tdc_details_2}}
\end{figure*}

Numerical simulation have shown that in the Ruderman-Sutherland model
pair creation is quasi-periodic and self-sustained.  We performed
simulations for different initial particle distributions and different
initial electric fields, different strengths of the magnetic field,
different radii of curvature of magnetic field lines and for different
pulsar periods.  After the initial burst of pair creation the cascade
zone always settled down to a quasi-periodic behavior qualitatively
similar for all physical parameters admitting pair creation.  The
system seems to forget initial conditions and after a short relaxation
(couple of flyby times) for given magnetic field, pulsar period and
the mean current $j_0$ its behavior is the same independent on the
initial configuration.

We describe here main properties of the cascade using as an example
the case with the mean current density equal to the Goldreich-Julian
\citep{GJ} current density, $j_0=\GJ{j}$.  For the mean current
density different from $\GJ{j}$ cascade behavior is qualitatively
similar.  In this example we consider a pulsar with period
$P-0.2$~sec, magnetic field $B=10^{12}$~G, radius of curvature of the
magnetic field lines $\rho_{\mbox{cur}}=10^6$~cm (such value for
$\rho_{\mbox{cur}}$ was used in \citep{Ruderman/Sutherland75}).  We
performed simulations for pure dipole magnetic field with
$\rho_{\mbox{cur}}\sim10^8$~cm too.  Qualitatively results do not
depend on the radius of curvature, but for smaller $\rho_{\mbox{cur}}$
calculations with the same numerical resolution can be done faster, as
the size of the gap with accelerating electric field is smaller.
Angular velocity of NS rotation is anti-parallel to the magnetic
moment of the star, so the Goldreich-Julian change density is
positive.

We describe below a whole cycle of pair formation and illustrate the
cascade development by a series of snapshots at several time moments
during one cycle with plots showing different physical quantities in
Figs.~\ref{fig:tdc_rho_timeseries},~\ref{fig:tdc_details},~\ref{fig:tdc_details_2}.
In these figures we present plots for a typical cycle of pair
formation taken from a long simulation where several such cycles were
observed.  The time on this figures is normalized to the flyby time of
the computational domain.  Time is counted from the start of a
particular simulation, so its absolute value has no physical meaning
-- only time intervals between the shots have physical meaning.

Changes in the charge density distribution gives the best overview of
what is going on in the discharge zone, because charge density
indirectly provides information about both the particle number density
and the electric field.  In Fig.~\ref{fig:tdc_rho_timeseries} we plot
the change density at equally spaced time interval during the
discharge cycle.  In the upper panel of that figure we present an
overview of the entire cycle, in the lower panel we plot snapshots of
the change density distribution at smaller time intervals for the most
interesting part of the discharge -- formation of a new plasma blob.
In Figs.~\ref{fig:tdc_details},~\ref{fig:tdc_details_2} we show
more detailed information about physical conditions in the discharge
zone: the number densities of electrons and positron, the accelerating
electric field, phase portraits of electrons, positrons and pair
producing photons.  On the phase portraits particles with positive
values of the 4-momentum $p$ are those which move from the NS,
particles with negative $p$ move toward the NS.

A typical cycle of the discharge starts with a vacuum gap forming
above the surface of the NS (snapshots with $t=8.9-9.5$). The
electric field there is very strong and charged particles entering the
gap are accelerated up to very high energies.  Plasma creation starts
close to the NS and is ignited by the gamma-rays emitted by electrons
flowing \emph{toward} the NS. These ``primary'' electrons have been
created in the previous bursts of pair formation. They leaks from the
tail of the plasma blob created in the previous cycle and enters the
gap from above. The newly created electrons and positrons are
accelerated by the strong electric field of the gap and start
producing high energy photons, many of which decay into pairs too
(snapshot with $t=9.667$ in
Figs.~\ref{fig:tdc_details},~\ref{fig:tdc_rho_timeseries}).  In
Fig.~\ref{fig:tdc_details},~\ref{fig:tdc_details_2} on the plots
showing the phase portrait for electrons ``primary'' electrons
accelerated in the gap can be seen as the particle population having
thin line-like form.  Particle populations scattered over the large
area beginning at the left end of the phase space plots represent
secondary particles.

While number density of the pair plasma remains less then the
Goldreich-Julian density, the electric field remains strong and
electrons and positrons are accelerated to energies high enough to
emit photons capable of producing pairs (see snapshots with
$t=9.667-9.8$).  When the number density become larger than the
Goldreich-Julian density, acceleration of particles ceases and
particles created after that moment do not emit pair-producing
photons.  Screening of the electric field begins near the NS surface,
where first pairs are formed, so the region of a very low plasma
density (the gap) with strong electric field detaches from the NS
surface and propagates into the magnetosphere (snapshots with
$t>9.933$).  From below this gap is limited by the blob of freshly
formed plasma, from above -- by the plasma created in the previous
burst of pair formation.  Particles from the previous bursts of pair
formation which have momentum directed toward the NS enters the gap
from above.  Electrons are accelerated toward the NS, positrons are
turned back by the strong electric field of the gap and move into the
magnetosphere.  Because of this reversal the gap upper boundary moves
noticeably slower than the speed of light.  The front of the blob with
the freshly created plasma consists of ultra-relativistic positrons
accelerated in then strong electric field, so it moves
relativistically and, therefore, the gap shrinks when the blob moves
into the magnetosphere.  Pairs are continuously injected into the
blob, because it practically co-moves with the pair-producing photons.
Eventually the front of the new blob catches the tail of the
previously created blob.  Therefore, the magnetosphere will be filled
with plasma and there will be no gaps in plasma spacial distribution
far from the polar cap.  When the blob has traveled some distance a
new gap starts forming, ignited by the electrons leaving the current
blob.

When the electric field is still strong large amplitude plasma
oscillations are excited in the newly formed plasma.  These
oscillations persists after the global accelerating field has been
screened.  Fluctuating electric field of the oscillations reverses
some electrons and positrons toward the NS.  These particles forms the
above mentioned tail of the plasma blob.  The electrons from that tail
will be the ``primary'' particles in the next burst of pair formation.

The height of the gap is $\sim2$ times larger than estimate given by
eq.~(\ref{eq:h_RS}).  The maximum particle energy is $\sim4$ higher
than given in \citep{Ruderman/Sutherland75} and particle energy
distributions is broad.

\section{Conclusions and open issues}

Main findings from the numerical simulations can be summarized as
follows.  The Ruderman-Sutherland cascade can operate for any positive
current density.  The cascade is self-sustained and discharges occur
quasi-periodically, the whole systems shows a sort of limit cycle
behavior.  As cascade properties do not depend on the initial
conditions in our simulations, they are uniquely set by the mean
current density $j_0$, the potential drop across the polar cap and the
curvature of the magnetic field lines.  Distribution of the pair
plasma in the magnetosphere is continuous, with no gaps, so the
parallel component of the electric field far from the polar cap should
be screened.  Thus, such cascade model agrees well with force-free
magnetosphere models
\citep[e.g.][]{CKF,Timokhin2006:MNRAS1,Spitkovsky:incl:06}.  Strong
plasma oscillations are excited in the plasma blob during its
formation.  This might have implication for generation of
radioemission.

While the general structure of the flow is evident from the performed
simulations, some questions remain unanswered. The most important one
is about the period of these discharge. It depends on the rate of
plasma leakage from the blob, i.e. how many particles are reversed
toward the NS.  The more particles leaks out, the later the next gap
forms. Due to continuous pair injection plasma density in the blob
increases enormously, and at some time the numerical scheme stops
resolving the Debye length of the plasma, and the damping of the
plasma oscillations cannot be calculated reliably.  At that time
results start depending on the numerical resolution.  Because of this
the blob cannot be followed for time interval long enough (and
traveled distances large enough) to get the repetition rate of the
cascade.  In the presented simulations the size of the simulation zone
is set such that the blob leaves calculation domain before the
numerical scheme fails to model it correctly.  So, while particle
energy distribution can be inferred from current simulations, pair
multiplicity and particle fluxes are known up to some numerical factor
which depends on the repetition rate of pair creation bursts.

More detailed description of the results as well as discussion of
their implication for pulsar physics will be given in a subsequent
publication.

\begin{acknowledgments}
I'm deeply indebted to Jonathan Arons for encouragement, continuous
support, and for innumerable exciting discussions which substantially
influenced my understanding of the problem. This work was supported by
NSF grant AST-0507813 and NASA grants NNG06GJI08G and NNX09AU05G.
\end{acknowledgments}

\bibliography{/home/atim/ARTICLES/Bibliographies/pulsars/pulsars_theory,/home/atim/ARTICLES/Bibliographies/pulsars/pulsars_obs,/home/atim/ARTICLES/Bibliographies/magnetars/magnetars,/home/atim/ARTICLES/Bibliographies/NumericalMethods/numerical_methods}

\begin{thebibliography}{14}
\expandafter\ifx\csname natexlab\endcsname\relax\def\natexlab#1{#1}\fi
\expandafter\ifx\csname bibnamefont\endcsname\relax
  \def\bibnamefont#1{#1}\fi
\expandafter\ifx\csname bibfnamefont\endcsname\relax
  \def\bibfnamefont#1{#1}\fi
\expandafter\ifx\csname citenamefont\endcsname\relax
  \def\citenamefont#1{#1}\fi
\expandafter\ifx\csname url\endcsname\relax
  \def\url#1{\texttt{#1}}\fi
\expandafter\ifx\csname urlprefix\endcsname\relax\def\urlprefix{URL }\fi
\providecommand{\bibinfo}[2]{#2}
\providecommand{\eprint}[2][]{\url{#2}}

\bibitem[{\citenamefont{{Arons} and {Scharlemann}}(1979)}]{Arons1979}
\bibinfo{author}{\bibfnamefont{J.}~\bibnamefont{{Arons}}} \bibnamefont{and}
  \bibinfo{author}{\bibfnamefont{E.~T.} \bibnamefont{{Scharlemann}}},
  \bibinfo{journal}{\apj} \textbf{\bibinfo{volume}{231}}, \bibinfo{pages}{854}
  (\bibinfo{year}{1979}).

\bibitem[{\citenamefont{{Daugherty} and {Harding}}(1982)}]{Daugherty/Harding82}
\bibinfo{author}{\bibfnamefont{J.~K.} \bibnamefont{{Daugherty}}}
  \bibnamefont{and} \bibinfo{author}{\bibfnamefont{A.~K.}
  \bibnamefont{{Harding}}}, \bibinfo{journal}{\apj}
  \textbf{\bibinfo{volume}{252}}, \bibinfo{pages}{337} (\bibinfo{year}{1982}).

\bibitem[{\citenamefont{{Muslimov} and {Tsygan}}(1992)}]{Muslimov/Tsygan92}
\bibinfo{author}{\bibfnamefont{A.~G.} \bibnamefont{{Muslimov}}}
  \bibnamefont{and} \bibinfo{author}{\bibfnamefont{A.~I.}
  \bibnamefont{{Tsygan}}}, \bibinfo{journal}{\mnras}
  \textbf{\bibinfo{volume}{255}}, \bibinfo{pages}{61} (\bibinfo{year}{1992}).

\bibitem[{\citenamefont{{Harding} and
  {Muslimov}}(2002)}]{Harding/Muslimov:heating_2::2002}
\bibinfo{author}{\bibfnamefont{A.~K.} \bibnamefont{{Harding}}}
  \bibnamefont{and} \bibinfo{author}{\bibfnamefont{A.~G.}
  \bibnamefont{{Muslimov}}}, \bibinfo{journal}{\apj}
  \textbf{\bibinfo{volume}{568}}, \bibinfo{pages}{862} (\bibinfo{year}{2002}).

\bibitem[{\citenamefont{{Hibschman} and
  {Arons}}(2001)}]{Hibschman/Arons:pair_multipl::2001}
\bibinfo{author}{\bibfnamefont{J.~A.} \bibnamefont{{Hibschman}}}
  \bibnamefont{and} \bibinfo{author}{\bibfnamefont{J.}~\bibnamefont{{Arons}}},
  \bibinfo{journal}{\apj} \textbf{\bibinfo{volume}{554}}, \bibinfo{pages}{624}
  (\bibinfo{year}{2001}).

\bibitem[{\citenamefont{{de Jager}}(2007)}]{deJager2007}
\bibinfo{author}{\bibfnamefont{O.~C.} \bibnamefont{{de Jager}}},
  \bibinfo{journal}{\apj} \textbf{\bibinfo{volume}{658}}, \bibinfo{pages}{1177}
  (\bibinfo{year}{2007}).

\bibitem[{\citenamefont{{Timokhin}}(2006)}]{Timokhin2006:MNRAS1}
\bibinfo{author}{\bibfnamefont{A.~N.} \bibnamefont{{Timokhin}}},
  \bibinfo{journal}{\mnras} \textbf{\bibinfo{volume}{368}},
  \bibinfo{pages}{1055} (\bibinfo{year}{2006}).

\bibitem[{\citenamefont{{Ruderman} and
  {Sutherland}}(1975)}]{Ruderman/Sutherland75}
\bibinfo{author}{\bibfnamefont{M.~A.} \bibnamefont{{Ruderman}}}
  \bibnamefont{and} \bibinfo{author}{\bibfnamefont{P.~G.}
  \bibnamefont{{Sutherland}}}, \bibinfo{journal}{\apj}
  \textbf{\bibinfo{volume}{196}}, \bibinfo{pages}{51} (\bibinfo{year}{1975}).

\bibitem[{\citenamefont{{Sturrock}}(1971)}]{Sturrock71}
\bibinfo{author}{\bibfnamefont{P.~A.} \bibnamefont{{Sturrock}}},
  \bibinfo{journal}{\apj} \textbf{\bibinfo{volume}{164}}, \bibinfo{pages}{529}
  (\bibinfo{year}{1971}).

\bibitem[{\citenamefont{{Al'Ber} et~al.}(1975)\citenamefont{{Al'Ber},
  {Krotova}, and {{E}idman}}}]{AlBer/Krotova:1975}
\bibinfo{author}{\bibfnamefont{Y.~I.} \bibnamefont{{Al'Ber}}},
  \bibinfo{author}{\bibfnamefont{Z.~N.} \bibnamefont{{Krotova}}},
  \bibnamefont{and} \bibinfo{author}{\bibfnamefont{V.~Y.}
  \bibnamefont{{{E}idman}}}, \bibinfo{journal}{Astrophysics}
  \textbf{\bibinfo{volume}{11}}, \bibinfo{pages}{189} (\bibinfo{year}{1975}).

\bibitem[{\citenamefont{{Levinson} et~al.}(2005)\citenamefont{{Levinson},
  {Melrose}, {Judge}, and {Luo}}}]{Levinson05}
\bibinfo{author}{\bibfnamefont{A.}~\bibnamefont{{Levinson}}},
  \bibinfo{author}{\bibfnamefont{D.}~\bibnamefont{{Melrose}}},
  \bibinfo{author}{\bibfnamefont{A.}~\bibnamefont{{Judge}}}, \bibnamefont{and}
  \bibinfo{author}{\bibfnamefont{Q.}~\bibnamefont{{Luo}}},
  \bibinfo{journal}{\apj} \textbf{\bibinfo{volume}{631}}, \bibinfo{pages}{456}
  (\bibinfo{year}{2005}).

\bibitem[{\citenamefont{{Goldreich} and {Julian}}(1969)}]{GJ}
\bibinfo{author}{\bibfnamefont{P.}~\bibnamefont{{Goldreich}}} \bibnamefont{and}
  \bibinfo{author}{\bibfnamefont{W.~H.} \bibnamefont{{Julian}}},
  \bibinfo{journal}{\apj} \textbf{\bibinfo{volume}{157}}, \bibinfo{pages}{869}
  (\bibinfo{year}{1969}).

\bibitem[{\citenamefont{{Contopoulos} et~al.}(1999)\citenamefont{{Contopoulos},
  {Kazanas}, and {Fendt}}}]{CKF}
\bibinfo{author}{\bibfnamefont{I.}~\bibnamefont{{Contopoulos}}},
  \bibinfo{author}{\bibfnamefont{D.}~\bibnamefont{{Kazanas}}},
  \bibnamefont{and} \bibinfo{author}{\bibfnamefont{C.}~\bibnamefont{{Fendt}}},
  \bibinfo{journal}{\apj} \textbf{\bibinfo{volume}{511}}, \bibinfo{pages}{351}
  (\bibinfo{year}{1999}).

\bibitem[{\citenamefont{{Spitkovsky}}(2006)}]{Spitkovsky:incl:06}
\bibinfo{author}{\bibfnamefont{A.}~\bibnamefont{{Spitkovsky}}},
  \bibinfo{journal}{\apjl} \textbf{\bibinfo{volume}{648}}, \bibinfo{pages}{L51}
  (\bibinfo{year}{2006}).

\end{thebibliography}

\end{document}